\documentclass[12pt]{article}
\usepackage{epsf}
\usepackage{cite}
\usepackage{axodraw}

\begin{document}

\newcommand{\TSOL}[1]{
\SetScale{#1}
\Vertex(0,0){8}
\Vertex(80,40){5}
\Vertex(80,-40){5}
\ArrowLine(-30,0)(0,0)
\ArrowLine(80,40)(100,40)
\ArrowLine(80,-40)(100,-40)
\ArrowLine(0,0)(80,40)
\ArrowLine(0,0)(80,-40)
\Line(80,-40)(80,40)
%
}
\newcommand{\TSTL}[1]{
\SetScale{#1}
\Vertex(0,0){8}
\Vertex(80,50){5}
\Vertex(80,-50){5}
\Vertex(80,0){5}
\ArrowLine(-30,0)(0,0)
\ArrowLine(80,50)(100,50)
\ArrowLine(80,-50)(100,-50)
\ArrowLine(0,0)(80,50)
\ArrowLine(0,0)(80,0)
\ArrowLine(0,0)(80,-50)
\Line(80,-50)(80,50)
%
}
\newcommand{\MIX}[1]{
\SetScale{#1}
\Vertex(0,0){8}
\Vertex(40,-20){5}
\ArrowLine(-30,0)(0,0)
\ArrowLine(0,0)(60,30)
\ArrowLine(0,0)(40,-20)
\ArrowLine(40,-20)(80,-40)
}
\newcommand{\RRM}[1]{
\SetScale{#1}
\Vertex(0,0){8}
\Vertex(40,-40){5}
\Vertex(70,-70){5}
\ArrowLine(-30,0)(0,0)
\ArrowLine(0,0)(40,40)
\ArrowLine(40,-40)(70,-70)
\ArrowArc(40,0)(40,180,270)
\ArrowArcn(0,-40)(40,90,0)
\ArrowArc(110,-70)(40,180,225)
\ArrowArcn(70,-110)(40,90,45)
}
\newcommand{\ttbs}{\char'134}
\newcommand{\AmS}{{\protect\the\textfont2
  A\kern-.1667em\lower.5ex\hbox{M}\kern-.125emS}}

\hyphenation{author another created financial paper re-commend-ed}



\begin{center}
{\LARGE\sf Loop Calculations and OZI Rule}\\[5mm]
{\sc V.~E.~Markushin}\\[2mm] 
{\it Paul Scherrer Institute, CH-5232 Villigen, Switzerland}\\[5mm]
{December 2, 1998}\\[5mm]
{\small Invited contribution at LEAP-98, Villasimius, September 7--12, 1998 \\ 
        (to be published in Nuclear Physics)}\\[5mm]
\end{center}

\begin{abstract}

  The role of two step mechanisms in the OZI rule violation in $N\bar{N}$
annihilation at low energies is reviewed.
  The full calculation of two-meson doorway mechanisms shows that
the off-mass-shell contributions are important.
  Moreover, for the $\phi\pi^0$ and $\phi\rho$ annihilation channels 
three-meson doorway contributions are non-negligible and improve agreement 
with the data.
  A significant enhancement of $\phi$ production is found in the
resonant rescattering mechanisms  $\bar{p}p\to\pi^+\pi^-
K\bar{K}\to\pi^+\pi^-\phi$ and $\bar{p}p\to\eta K\bar{K}\to\eta\phi$.

\end{abstract}

   \section{Introduction}

   The $\phi$ production in low energy $p\bar{p}$ annihilation is
expected to be suppressed on the tree level according to the
Okubo-Zweig-Iizuka (OZI) rule \cite{Ok63,Zw64,Ii66} because the $\phi$
meson is nearly a pure $s\bar{s}$ state.  The $\phi$ production via
$\phi\omega$ mixing (Fig.\ref{FigMDWM}a) is proportional to a small
deviation of the vector octet mixing angle $\theta_V$ from the ideal one
$\theta_i$ ($\omega_{\phi\omega}=\sin(\theta_V-\theta_i)\approx 0.065$)
\cite{PDG}.  The corresponding amplitude $T^{\mathrm OZI}$ determines
the expected scale of $\phi$ production.  The reactions where this
level is significantly exceeded (OZI rule violation) are of big interest
for the studies of annihilation mechanisms and nucleon structure.

   The experiments performed at LEAR (see
\cite{AST91,OBEL1,OBEL2,OBEL95,CBC95,CBC97,JETSET95a,OBEL96,OBEL98,OBEL98a,Sa98}
and references therein)
have demonstrated, in agreement with earlier findings \cite{preLEAR},
that the OZI rule is strongly violated in several channels in $p\bar{p}$ annihilation
at low energy.
   To explain these results two different theoretical approaches have
been used.  One approach \cite{LZL93,LLZ94,LL94,BL94,MHS,BL95,BL95a,GLMR,AK96}
is based on higher order processes beyond the tree level (rescattering mechanisms),
the second one assumes that the nucleon has a significant intrinsic
$s\bar{s}$ component \cite{EGK89,EKKS95}.
   This paper, supplementing the earlier reviews
\cite{Lo95,Sa95,Zou96,Ma97,Wi97}, focuses on recent studies of the
rescattering mechanisms in $p\bar{p}$ annihilation.

   \section{Two-step mechanisms}

\begin{figure}
\centering
\mbox{(a) \hspace*{30mm} (b) \hspace*{30mm} (c) \hspace*{10mm} \ } \\
\mbox{

   \mbox{
      \begin{picture}(100,60)(-25,-30)
      \MIX{0.5}
      \Text(-10,6)[b]{$p\bar{p}$}
      \Text(40,20)[l]{$X$}
      \Text(42,-20)[l]{$\phi$}
      \Text(5,-8)[r]{$\omega$}
      \Text(15,-12)[t]{$\epsilon_{\omega\phi}$}
      \end{picture}
   }

   \mbox{     
      \begin{picture}(100,50)(-25,-28)
      \TSOL{0.5}
      \Text(-10,6)[r]{$p\bar{p}$}
      \Text(60,20)[l]{$X$}
      \Text(60,-20)[l]{$\phi$}
      \Text(20,16)[b]{$2$}
      \Text(20,-17)[t]{$1$}
      \Text(43,0)[l]{$3$}
      \end{picture}
   }

   \mbox{     
       \begin{picture}(100,60)(-25,-30)
       \TSTL{0.5}
       \Text(-10,6)[b]{$p\bar{p}$}
       \Text(55,25)[l]{$X$}
       \Text(55,-25)[l]{$\phi$}
       \Text(22,19)[b]{$2$}
       \Text(22,2)[b]{$3$}
       \Text(22,-20)[t]{$1$}
       \Text(41,10)[l]{$5$}
       \Text(41,-10)[l]{$4$}
       \end{picture}
   }
}\\[0mm]
\caption{\small\label{FigMDWM}
(a) The $\phi\omega$ mixing and
the two-step processes for $\phi$ production:
(a) two-meson doorway mechanism with intermediate state $1\;2$,
(b) three-meson doorway mechanism with intermediate state $1\;2\;3$.
}
\end{figure}
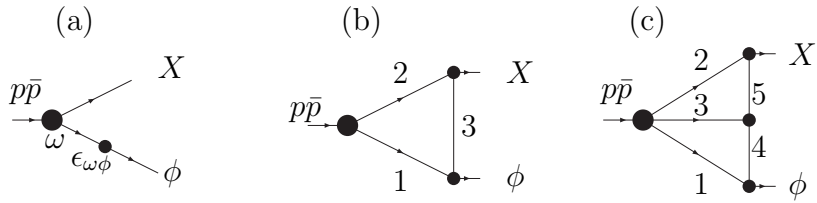

\begin{table}
\caption{\small\label{TabTSM}
The contributions from $\phi\omega$ mixing, from two-meson doorway 
and from three-meson doorway mechanisms to amplitude $T$ and 
branching ratio $BR$  
for the reactions
$(p\bar{p})_{^3S_1}\to\phi\pi^0$ and
$(p\bar{p})_{^1S_0}\to\phi\rho^0$ at rest \protect\cite{GLMR,RLM98}.
The intermediate states and the mesons exchanged are indicated
according to Figs.\protect\ref{FigMDWM}b,c.
   The amplitude $T^{\mathrm OZI}$ corresponds to $\phi\omega$
mixing with a point-like $p\bar{p}\to\phi X$ vertex; the effect of a
dipole form factor in the annihilation vertex with cut-off parameter
$\Lambda$ is shown for the $\phi\rho$ case.
}
\vspace*{3mm}
\begin{tabular}{lllll}
\hline {\rule{0mm}{5mm}}
  reaction & \multicolumn{2}{l}{mechanism}
  & $|T/T^{\mathrm OZI}|$
  & $BR(\phi X)\cdot 10^4$ \\[1mm]
\hline
  \fbox{$p\bar{p}\to\phi\pi^0$}
 & Fig.\ref{FigMDWM}a & $\phi\omega$ mixing    &  1.          &  0.15  \\
\cline{2-5}
 & Fig.\ref{FigMDWM}b
        & $\rho\rho$ ($\pi$)            &  $0.6-2.6$   & $0.05-1.0$ \\
 &      & $K\bar{K}^*+\bar{K}K^*$ ($K$) &  $2.0-3.3$   & $0.6-1.6$  \\
 &      & $K\bar{K}$ ($K^*$)            &  $<0.5$      & $<0.03$  \\
 &      & $K^*\bar{K}^*$ ($K$)          &  $<0.7$      & $<0.08$  \\
\cline{2-5}
 & Fig.\ref{FigMDWM}c
        & $\pi\rho\pi$ ($\rho,\omega$)  &  $\sim 2.0$  & $\sim 0.6$ \\
 &      & $\rho\pi\pi$ ($\pi,\rho$)     &  $\sim 1.2$  & $\sim 0.2$ \\
\cline{2-5}
 & \multicolumn{2}{l}{experiment \cite{AST91}}
                                        &  $5.2$       & $4.0 \pm 0.8$ \\
 & \multicolumn{2}{l}{experiment \cite{CBC95,CBC97}}
                                        &  $6.7$       & $6.5 \pm 0.6$ \\
 & \multicolumn{2}{l}{experiment \cite{OBEL98a}}
                                        &  $7.2$       & $7.6 \pm 0.6$
\\ \cline{2-5}
 & & & & \\[-2mm]
\hline
  \fbox{$p\bar{p}\to\phi\rho^0$}
  & Fig.\ref{FigMDWM}a & $\phi\omega$ mixing   &  1.          &  0.14  \\
  &                    & $\Lambda=0.4\;$GeV     &  1.8         &  0.47
\\
\cline{2-5}
 & Fig.\ref{FigMDWM}b
        & $\omega\rho$ ($\pi$)          &  $<0.6$      & $<0.05$ \\
 &      & $K\bar{K}^*+\bar{K}K^*$ ($K$) &  $<0.3$      & $<0.01$  \\
\cline{2-5}
 & Fig.\ref{FigMDWM}c
        & $\pi\omega\pi+\pi\pi\omega$ ($\rho,\pi$)
                                        &  $\sim 1.1$  & $\sim 0.17$ \\
 &      & $\pi\pi\pi$ ($\rho,\pi$)   &  $\sim 0.6$  & $\sim 0.06$ \\
\cline{2-5}
 & \multicolumn{2}{l}{experiment \cite{AST91}}
                                        &  $4.9$       & $3.4 \pm 1.0$ \\
\cline{2-5}
 & & & & \\[-2mm]
\hline

\end{tabular}
\end{table}

  Two-step processes with ordinary hadrons in intermediate states
are known to be a potential source of the OZI-rule breaking
(see \cite{Li84,LZ96}).
   The simplest two-step process in $p\bar{p}$ annihilation is the
two-meson doorway mechanism shown in
Fig.\ref{FigMDWM}b.  Its role has been studied for various final states
containing $\phi$ mesons
\cite{LZL93,LLZ94,LL94,BL94,MHS,BL95,BL95a,GLMR,AK96}.
   In the $\phi\pi$ and $\phi\phi$ channels presenting the
most dramatic hadronic violation of the OZI rule,
two-step mechanisms predict cross sections comparable with the
experimental data.
  The full calculation of two-meson doorway mechanisms \cite{GLMR,RLM98}
shows that off-mass-shell contributions are of similar size as
the unitarity approximation.
%
%
   The contributions from different intermediate states in the
reaction $p\bar{p}\to\phi\pi^0$ are shown in Table~\ref{TabTSM}.
   A constructive interference between the two-meson doorway
contributions from the $K\bar{K}^*$ and $\rho\rho$ intermediate states
\cite{GLMR} is already sufficient to explain the measured branching
ratio $BR(\phi\pi^0)$ \cite{AST91,CBC95,CBC97,OBEL98a}.

   However, for a number of reactions, including
$p\bar{p}\to\phi\rho,\;\phi\pi\pi,\;\phi\omega$, no significant
contributions from the two-meson doorway mechanisms have been found
\cite{BL94}.
   Since the annihilation into two non-strange mesons is less
probable than into three mesons, three mesons are natural candidates
for intermediate states leading to sizeable OZI-rule violation.
   Three meson doorway mechanisms for the reactions
$(p\bar{p})_{^3S_1}\to\phi\pi^0,\phi\eta$ and
$(p\bar{p})_{^1S_0}\to\phi\rho,\phi\omega$
at rest have been considered in \cite{RLM98} using the unitarity
approximation (only the absorptive part of the diagrams in
Fig.\ref{FigMDWM}c was calculated).
   For the $\phi\pi$ and $\phi\rho$ channels,
three particle intermediate states
were found to be important, as shown in Table~\ref{TabTSM}.

   For all the channels considered, a consistent explanation of large
and small OZI-rule violations emerges.  In particular, the contribution of
the doorway mechanisms relative to tree level $\phi\omega$ mixing
is largest in the $\phi\pi^0$ channel where the most
dramatic OZI rule violation in hadronic channels is observed.
   As required, the relative strength of the doorway mechanisms is
smaller in channels with less dramatic breaking of the OZI rule.
   Furthermore, a consistency check was done by calculating
the two-step mechanisms for annihilation into nonstrange mesons; the
rescattering terms were found to be smaller than the tree-level terms
strengthening the case for the consistency of the rescattering
expansion \cite{RLM98}.

   While the rescattering mechanisms considered can explain the discussed
OZI-rule violating reactions, exact predictions are often hindered by
the lack of independent information about the relative phases of the
competing contributions from different intermediate states.  Lipkin
cancellations \cite{Li84,LZ96} seem to be unlikely in $p\bar{p}$
annihilation because the standard arguments based on the symmetry of
intermediate states (like in the case of the vector meson octet mixing)
do not apply.
   As an example where interference effects are important, we recall that the
OZI rule violation in the $p\bar{p}\to\phi\gamma$ at rest finds its
natural explanation in the framework of the vector meson dominance where
a constructive interference between the intermediate states $\phi\omega$
and $\phi\rho$ is required, contrary to the case of annihilation into
$\omega\gamma$ channel where the interference between the $\omega\omega$
and $\omega\rho$ is destructive
(see discussion in \cite{Ma97} and references therein).

   \section{Resonant rescattering mechanism}

\begin{figure}
\begin{center}
(a) \hspace{60mm} (b) \\[-5mm]
\mbox{
\mbox{\epsfxsize=70mm \epsffile{ozimkkthexp.epsf}}
\mbox{\epsfxsize=70mm \epsffile{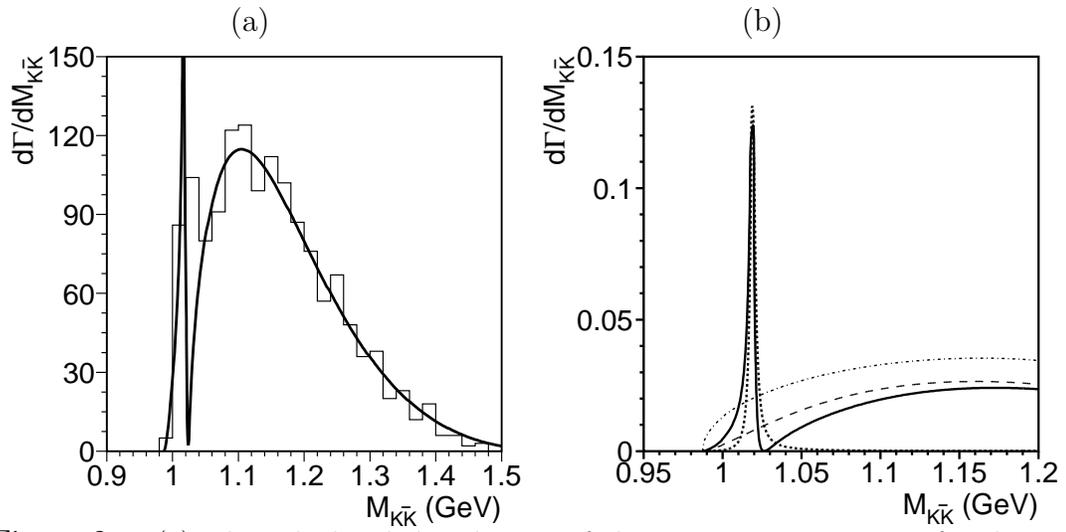}}
}
\end{center}
\vspace*{-10mm}
\caption{\small\label{FigKKmass}
(a) The calculated distribution of the invariant mass $m_{K\bar{K}}$ for
the reaction
$\bar{p}p(^3S_1)\to K\bar{K}\pi^+\pi^-$
for $\beta=2$ in comparison with the experimental data
\protect\cite{Ba66}.
(b) The calculated distribution of the invariant mass $m_{K\bar{K}}$ for
the reaction
$\bar{p}p\to K\bar{K}\eta$ for $\beta=4$.
The full result is shown by the solid line,
the pure resonance term by the dotted line,
the plane wave approximation by the dashed line,
and the phase space distribution by the dashed-dotted line,
respectively.
}
\end{figure}

   A further type of rescattering mechanisms proceeding via resonant final
state interaction in the $K\bar{K}$ was suggested recently in
\cite{ML98} where the reaction $p\bar{p}\to\phi\pi\pi$ was studied in
detail.
The total annihilation amplitude $T$ has the form:
\begin{eqnarray}
  T & = &
     \raisebox{-20mm}{
     \begin{picture}(220,90)(-150,-60)
     \RRM{0.5}
     \Vertex(-200,0){8}
     \ArrowLine(-200,0)(-150,45)
     \ArrowLine(-200,0)(-150,15)
     \ArrowLine(-200,0)(-150,-15)
     \ArrowLine(-200,0)(-150,-45)
     \ArrowLine(-230,0)(-200,0)
     \Text(-115,0)[r]{$p\bar{p}$}
     \Text(-73,23)[l]{$\pi^+$}
     \Text(-73,8)[l]{$\pi^-$}
     \Text(-73,-8)[l]{$K$}
     \Text(-73,-23)[l]{$\bar{K}$}
     \Text(-43,0)[c]{$+$}
     \ArrowLine(0,0)(40,20)
     \Text(-15,0)[r]{$p\bar{p}$}
     \Text(31,-25)[c]{$\phi$}
     \Text(25,23)[c]{$\pi^+$}
     \Text(25,10)[c]{$\pi^-$}
     \Text(17,-5)[l]{$K$}
     \Text(0,-17)[r]{$\bar{K}$}
     \Text(53,-35)[l]{$K$}
     \Text(35,-53)[r]{$\bar{K}$}
     \end{picture}
     }
= g_a A(M_{K\bar{K}})
\label{EqRRM}
\end{eqnarray}
where $g_a$ is the annihilation vertex in the plane wave approximation
and $A(s_{K\bar{K}})$ is the enhancement factor resulting from the
final state interaction in the $K\bar{K}$ system with the invariant
mass $M_{K\bar{K}}$:
\begin{eqnarray}
   A(s_{K\bar{K}}) & = &
   \frac{M_{K\bar{K}}^2-m_{\phi}^2-m_{\phi}\beta\Gamma_{\phi}}
        {M_{K\bar{K}}^2-m_{\phi}^2+im_{\phi}\Gamma_{\phi}}
\end{eqnarray}
Here $m_{\phi}$ and $\Gamma_{\phi}$ are the mass and width of the
$\phi$ meson and $\beta$ is the ratio of the real to imaginary part of the
$K\bar{K}$ loop in the second (rescattering) diagram in Eq.(\ref{EqRRM}).
   Note that the absorptive part of the rescattering amplitude, which
corresponds to $K$ and $\bar{K}$ on their mass shell (the unitarity
approximation), can be expressed through the corresponding vertex
functions in the physical region and is therefore well constrained
by experimental data. Thus the model has only one parameter,
$\beta$, which depends on the loop regularization. As shown
in \cite{ML98}, the off-shell contributions are important
for typical values $\beta>2$.

   The calculated distribution of the invariant mass $m_{K\bar{K}}$ for
the reaction $p\bar{p}\to K\bar{K}\pi^+\pi^-$ is shown in
Fig.\ref{FigKKmass}a in comparison with the experimental data.
The predicted branching ratio
$BR(p\bar{p}(^3S_1)\to\phi\pi^+\pi^-)\geq 3\cdot 10^{-4}$ for
$\beta\geq 2$ agrees well with the OBELIX
$BR((p\bar{p})_{\mathrm liq}\to\phi\pi^+\pi^-)=3.5(4)\cdot 10^{-4}$
\cite{OBEL96},
the ASTERIX analysis
$BR(p\bar{p}(S)\to\phi\pi^+\pi^-)=4.7(11)\cdot 10^{-4}$ \cite{AST91}
as well as with the values obtained from the
old bubble chamber measurements
$BR(p\bar{p}(S)\to\phi\pi^+\pi^-)=5.4(9)\cdot 10^{-4}$ \cite{Ba66}
(see \cite{ML98} for details)
and $BR(p\bar{p}(S)\to\phi\pi^+\pi^-)=4.6(9)\cdot 10^{-4}$ \cite{Bi69}.

   The theoretical prediction for the invariant mass $m_{K\bar{K}}$ in
the reaction $p\bar{p}(^3S_1)\to K\bar{K}\eta$ is shown for
$\beta=4$ in Fig.\ref{FigKKmass}b.
The relative strength of the $\phi$ peak depends sensitively on the
corresponding ratio $\beta$ which reflects the properties of the
$p\bar{p}\to\ K\bar{K}\eta$ amplitude.
   Its dependence on the initial $p\bar{p}$ state can be a reason for
the strong difference between the branching ratios
$BR(p\bar{p}(L)\to\phi\eta)$ for the $S$ and $P$-states
seen in recent OBELIX data \cite{OBEL98}.

   The interference of the resonant term with the nonresonant background
is essential for the correct analytical properties of the total
amplitude and leads to a characteristic peak-dip structure of the
invariant $K\bar{K}$ mass distribution, see Fig.\ref{FigKKmass}.
   Another interesting feature of the resonant rescattering mechanism
is that the $\phi$ production decreases with
increasing beam energy as the fraction of phase space favorable for
resonance formation gets smaller.  This is in agreement with the general
trend observed for the OZI rule violation in nucleon-antinucleon
annihilation.

   \section{Conclusion}

   Systematic calculations of the OZI rule violation due to two-step
mechanisms in the reactions
$p\bar{p}\;\to\; \phi\pi^0,\; \phi\rho^0,\; \phi\pi^+\pi^-,\; \phi\eta$
at rest are now available.
   In the $\phi\pi^0$ channel, the two-meson doorway mechanisms with
the $K\bar{K}^*$ and $\rho\rho$ intermediate states are sufficient,
assuming constructive interference, to explain the observed dramatic
violation of the OZI rule.
   In this case a sizeable contribution is also expected from the
three-meson doorway mechanisms.
   Three-meson doorway mechanisms are also found to be important in the
$\phi\rho$ channel which previously could not be explained by the
two-meson doorway mechanisms alone.
   The characteristic scale of the doorway mechanisms is well determined by the
experimental information.
The contributions from the doorway mechanisms are large for the channels
with dramatic violation of the OZI rule  and small for the channels
with moderate violation.
   Therefore, a self-consistent picture of $\phi$ production in
$p\bar{p}$ annihilation emerges.

  A resonant $K\bar{K}$ rescattering mechanism (final state interaction)
is important for $\phi$ production in the $p\bar{p}$ annihilation at
low energy.
  The contribution of this mechanism to the $\phi\pi\pi$ channel agrees
with experiment and is a good candidate for the explanation of the
annihilation into the $\phi\eta$ channel as well.  The $\phi$ production
via the resonant rescattering mechanism decreases with increasing
total energy in agreement with the observed general trend.

   \section{Acknowledgements}

The author is grateful to M.P.~Locher and S.~von~Rotz for fruitful
collaboration and to M.G.~Sapozhnikov for stimulating discussions.


\end{document}